\documentclass{aastex62}

\usepackage{threeparttable}
\usepackage{multirow}
\usepackage{booktabs}
\usepackage{lineno}

\graphicspath{{Figs/}}

\received{}
\revised{}
\accepted{}
\submitjournal{ApJL}

\shorttitle{Effect of Stellar Rotation Rates on the Characteristics of Energetic Particle Events}
\shortauthors{Fu et al.}

\begin{document}
\title{Effect of Star Rotation Rates on the Characteristics of Energetic Particle Events}

\correspondingauthor{Gang Li}
\email{gang.li@uah.edu}

\author[0000-0003-4245-3107]{Shuai Fu}
\affiliation{Institute of Space Weather, Nanjing University of Information Science and Technology, Nanjing, 210044, People's Republic of China}
\affiliation{Department of Space Science and CSPAR, University of Alabama in Huntsville, Huntsville, AL, USA}

\author{Yong Jiang}
\affiliation{Institute of Space Weather, Nanjing University of Information Science and Technology, Nanjing, 210044, People's Republic of China}

\author{Vladimir Airapetian}
\affiliation{ NASA Goddard Space Flight Center, MD, USA}

\author{Junxiang Hu}
\affiliation{Department of Space Science and CSPAR, University of Alabama in Huntsville, Huntsville, AL, USA}

\author[0000-0003-4695-8866]{Gang Li}
\affiliation{Department of Space Science and CSPAR, University of Alabama in Huntsville, Huntsville, AL, USA}

\author[0000-0002-4642-6192]{Gary Zank}
\affiliation{Department of Space Science and CSPAR, University of Alabama in Huntsville, Huntsville, AL, USA}

\begin{abstract}
 Recent detection of superflares on solar-type stars by Kepler mission raised a possibility that they can be associated with energetic coronal mass ejections (CMEs) and energetic particle events (SEPs). These space weather events can impact habitability of exoplanets around these stars. Here we use the improved Particle Acceleration and Transport in the Heliosphere (iPATH) model, to model the time intensity profile and spectrum of SEPs accelerated at CME-driven shocks from stars of different ages traced by their rotation rates.
We consider a solar-like (G-type) star with  $6$ different rotation rates varying from $0.5 \Omega_{\sun}$ to $3.0 \Omega_{\sun}$.
In all $6$ cases, a fast CME is launched with the same speed of $\sim 1500$ km/sec and the resulting time intensity profiles  at $3$ locations and and energy spectra at $5$ locations at 1 AU are obtained. The maximum particle energy at the shock front as a function of $r$ is also shown. Our results suggest that within $0.8$ AU the maximum particle energy at the shock front increases with the rotation rate of the star. However, event integrated spectra for the five selected locations along the CME path show complicated patterns. 
This is because the Parker magnetic field for rapidly rotating stars is more tightly winded. Our results can be used in estimating the radiation environments of
terrestrial-type exoplanets around solar-type stars.
\end{abstract}

\keywords{shock waves --- Sun: coronal mass ejections (CMEs) --- stars: rotation}

\section{Introduction} \label{sec:intro}
White-light solar flares and associated Coronal Mass Ejections (CMEs) are the two most energetic
components of space weather affecting the magnetospheric and ionospheric environments of our planet.
The energy released in a large eruptive process manifested in an energetic flare and CME can reach up to a few times of $10^{32}$ erg and are frequently associated with formation of energetic particles accelerated up to GeV per nucleon  (e.g., \citealp{Emslie2004,Mewaldt2006,Zank2000}), producing the so called Solar Energetic Particle (SEP) events.
In large SEP events CMEs and flares often occur together. When the fast CME propagates out, it drives a shock wave and particles are accelerated at the shock front via the diffusive shock acceleration (DSA) mechanism. Because the shocks are spatially extended and the acceleration often lasts over an extended period, accelerated particles can be observed at multiple locations that are longitudinally well separated and the time intensity profile of these events are ``gradual" in nature. SEPs are commonly regarded as the No.1 space hazard for astronauts and electric instruments onboard spacecraft \citep{Feynman2000}.

Modelling SEP events have been taken by a number of authors (e.g., \citealp{Kallenrode1997, Kota2000, Kozarev2010, Luhmann2007, Luhmann2010,Vainio2007}).
One particular model is the Particle Acceleration and Transport in the Heliosphere (PATH) model. The model was originally developed by \citet{Zank2000} who adopted an onion shell model of the CME-driven shock from which the acceleration and diffusion of energetic particles are followed.  The model was improved by \citet{Rice2003} who considered shocks with arbitrary strength and by \citet{Li2003} who extended the original PATH model by adding a transport module investigating particle propagation using a Monte-Carlo approach.  \citet{Li2005} further extended the model to include the acceleration and transport of energetic heavy ions. These earlier versions of the PATH model are all 1D. Realizing that the shock obliquity is important in deciding the maximum energy at the shock front, \citet{Li2012} attempted to include shock obliquity in the PATH model. However, in the work of \citet{Li2012}, the shock obliquity is treated as a fixed parameter and does not vary as the shock propagates out.
It is therefore still an intrinsic 1D model.

Extending the PATH to a 2D model was done by
\citet{Hu2017}. The newer version of the model is named iPATH for improved PATH model. Comparing to the original PATH model, iPATH has the capability of simultaneously simulating energetic particle time intensity profiles and spectra at different locations. This is important for understanding observations made at multiple spacecraft of the same event (e.g., \citealp{Reames1996,Richardson2014,Gomez2015}).
\citet{Hu2018} performed a simulation showing multiple spacecraft observations of the same events. These locations differ in longitudes and heliocentric distances.

Here we apply iPATH to model SEP events forming in the environments of solar-type stars.
Young solar-type stars detected by Kepler mission are fast rotators with rotation periods of 3-5 days \citep{Guedel2007}.  A significant fraction of these stars (including F-, G-, K-, and M- stars) show superflare events \citep{Maehara2012}. These superflares reach energy up to $10^{35}$ ergs and are associated with large starspots occupying up to a few percent of the stellar surface. Solar CMEs are often associated with energetic flares. 
The relation of fast CMEs can be extended to large superflares observed on young solar-type stars \citep{Gomez2018}. As in the case of solar CMEs, particles can be accelerated to very high energies at the front of these shock waves. These high energy particles are a major concern of exoplanetary space weather, a subject which has emerged as an important subfield of exoplanetary science and it has a crucial impact on exoplanetary habitability \citep{Airapetian2016,Airapetian2019,Airapetian2017b}.

The rotation rate of solar-type stars depends of the stellar age varying a few to a few tens of days. As stars age, they lose angular momentum via magnetized winds \citep{WeberDavis1967} and coronal mass ejections \citep{Aarnio2011}. Consequently, their rotation rates decrease over time. It is important to investigate how energetic particle environments vary in solar-like stars of various ages and characterize their effects on exoplanets within close-in habitable zones (HZs) around active stars \citep{Airapetian2017}. However, such a study requires the knowledge of many stellar space weather environmental parameters. Besides the stellar rotation rate, one would also need the stellar wind speed, the stellar magnetic field, and the CME speed. Here we focus only on the effect of stellar rotation rate on the resulting energetic particle characteristics and defer such a study to a later paper.

 %

In this study, we apply iPATH model as described in \citep{Hu2017}.
Here we emphasize the importance of shock geometry on the particle acceleration process. The maximum particle energy at the shock front depends on the total diffusion coefficient $\kappa$ (see equation~(4)) 
Depending on the rotation rate of the star, the upstream magnetic field can assume different spiral pattern. A faster rotating star will have its spiral magnetic field more winded, while a slower rotating star will have a more radially oriented magnetic field. Consequently, shock geometry can vary significantly for stars of different rotation rates. Therefore one expects that the shock acceleration process will also be different.


Because the geometry of the Parker spiral field strongly depends on the rotation rate, therefore the geometry of the shock also strongly depends on
the star rotation rate.  In this letter, we use iPATH model to simulate the properties of SEP events from six scenarios of stellar rotation rates relevant to young to old solar-like stars.

\section{Model setup} \label{sec:model}
\begin{figure}[htb]
   \centering
   \includegraphics[width=14.2cm, angle=0]{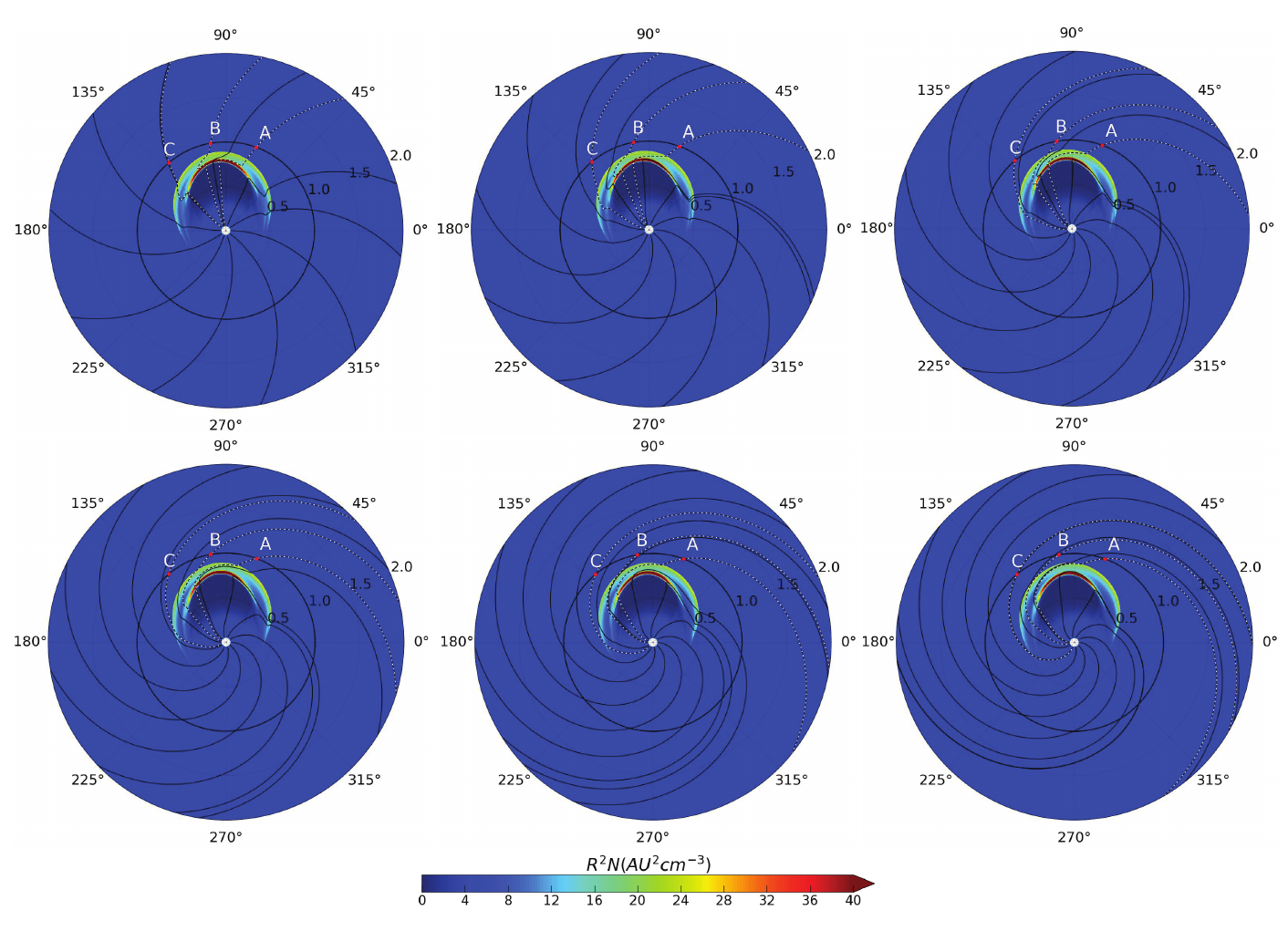}
   \caption{The configuration for the CME-driven shock under six star rotation scenarios at $t=30.2$ hours after CME eruption. The shock is initiated with the nose toward $\phi = 100^\circ$. {The top row, from left to right, are for $\Omega=0.5\Omega_{\sun}$ (case I), $1.0\Omega_{\sun}$ (case II),
     and $1.5\Omega_{\sun}$ (case III),
     respectively;  the bottom row,   from left to right, are for $\Omega=2.0\Omega_{\sun}$ (case IV), $2.5\Omega_{\sun}$ (case V), and $3.0\Omega_{\sun}$ (case VI),
     respectively. In the following case II is also referred to as the base case.}
     The color scheme is for the normalized density $nr^2$. The reference points labeled as A, B, and C locate at longitudes of $70^\circ$, $100^\circ$,
     and $130^\circ$, respectively. }
   \label{fig:fig1}
   \end{figure}

Denote $\Omega_{\sun}$ as the solar rotation rate, we examine six different runs with different rotation rates of $0.5\Omega_{\sun}$, $1.0\Omega_{\sun}$, and $1.5\Omega_{\sun}$, $2.0\Omega_{\sun}$, $2.5\Omega_{\sun}$, and $3.0\Omega_{\sun}$, respectively.
We model a steady background stellar wind environment in a spherical coordinate system with the background interplanetary magnetic field (IMF) given by a Parker spiral,
\begin{center}
\begin{equation}
B_{r}=B_{0}(\frac{R_{0}}{r})^2;B_{\phi}=B_{r}(\frac{{\Omega}r}{u_{sw}})(r\gg R_{0})\label{eq1}.
 \end{equation}
\end{center}
          {In the above, $B_r$ and $B_\phi$ are the radial and aziamuthal components of the star wind magnetic field at a heliocentric distance $r$, respectively. $\Omega$ is the star's rotation speed; $u_{sw}$ is the star wind speed;}
$R_0=0.05$ AU is the inner boundary. We carry out our simulations on a 2D domain (in the ecliptic plane) with 2000$\times$360 grid points covering a  radial range from $R_0$ to $2.0$ AU and a longitudinal range from $0^\circ$ to $360^\circ$. {The reference location at $0^{\circ}$ is arbitrarily chosen. The observer locations are specified relative to this reference location. To investigate the impact of magnetic field geometry on SEP properties, we keep the total magnetic field strength and the stellar wind speed at 1 AU as $4$ nT and $440$ km/s respectively.
  In future follow-up studies, we will apply a self-consistent treatment of the wind speed and the magnetic field which are derived for each stellar age scenario specified by the stellar rotation rate. }  {In all $6$ cases, we set proton number density and proton temperature to be $3800$/cm$^{3}$ and $3.8$ MK at the inner boundary, which yield a number density of
   $5$ cm$^{-3}$ and a temperature of $0.04$ MK at 1 AU, respectively.}
          To simulate a CME-driven shock structure, we perturb the star wind parameters in the inner boundary ($0.05$  AU, $\sim$10$R_\sun$) centering at longitude $\phi_c=100^\circ$.
At the center of the CME (corresponding to $\phi_c$), the star wind number density and temperature are increased by a factor of $4$ and $1.33$ from the ambient values and last $1$ hour. The magnetic field is not changed at the inner boundary. {The initial CME speed profile is taken as a
  Gaussian form. Denote the speed at longitude $\phi$ as $v(\phi)$ we have,}
  \begin{equation}
   v(\phi) = v_0 exp(-(\phi-\phi_c)^2/\sigma^2),
   \end{equation}
   {where $v_0=1469$ km/s is the speed at the center and $\sigma=66.6^\circ$ is the variance of the Gaussian distribution.}
    Figure~\ref{fig:fig1} is a snapshot of the CME-driven shock for the six different scenarios at $t=30.2$ hours. As can be seen from the figure, the faster the star rotates, the more twisted the IMF becomes. Points labeled as A, B, and C represent three observers locating at 1 AU with longitudes $70^\circ$, $100^\circ$, and $130^\circ$, respectively. The most important information from Figure~\ref{fig:fig1} is that the magnetic connections to the shock front at these three locations
      vary considerably. Our choices of A, B, and C are such that the shock nose passes point B, and the two flanks pass A and C.
    The iPATH model used in \cite{Hu2017,Hu2018} terminates at the shock arrival time, but the newly updated version is capable of simulating the shock arrival
    and downstream of the shock, therefore including the energetic storm particle (ESP) phase.

\section{Simulation results and Discussion} \label{sec:intro}

\begin{figure}[htb]
   \centering
   \includegraphics[width=15cm, angle=0]{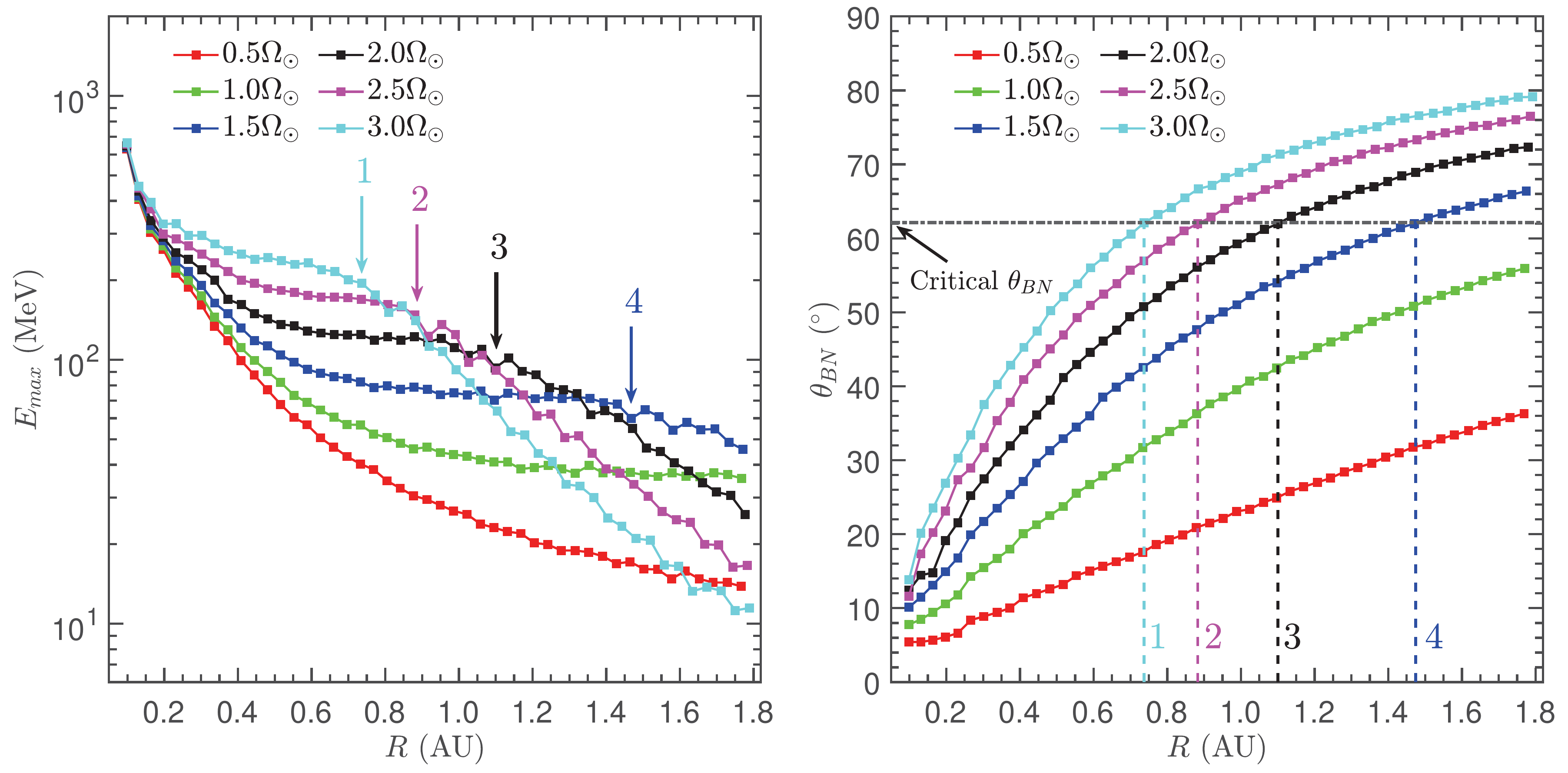}
   \caption{The maximum particle energy (left panel) and the shock obliquity angle (right panel) near the shock nose as a function of heliocentric distance $r$. The arrow (left panel) and the vertical dashed line (right panel) denote the position where the maximum energy curve falls down substantially. The horizontal dash-dotted line in the right panel shows the critical value of $\theta_{BN}$.
   }
   \label{fig:fig2}
   \end{figure}

Figure \ref{fig:fig2} plots the maximum particle energy and the shock obliquity angle near shock nose as a function of heliocentric distance.
The maximum particle energy is calculated by balancing the shock dynamic timescale $t_{dyn}$ with the particle acceleration timescale $t_{acc}$ (e.g., \citealp{Drury1983,Zank2000}),

\begin{equation}
t_{dyn}=\int_{p_0}^{p_{max}}\frac{3s}{s-1}\frac{\kappa}{U_{up}^{2}}\frac{1}{p}dp.\label{tdyn}
\end{equation}
{In the above, $p_0$ is the particle injection momentum, $p_{max}$ is the particle maximum momentum, $s$ is the
  shock compression ratio, $\kappa$ is particle's total diffusion coefficient at the shock, $U_{up}$ is the upstream plasma
  speed as seen in the shock frame. The total diffusion coefficient $\kappa$ is related to the parallel diffusion coefficient
  $\kappa_{||}$ and the perpendicular diffusion coefficient $\kappa_{\perp}$ and the shock obliquity angle $\theta_{BN}$ through,
\begin{equation}
  \kappa = \kappa_{||} cos^2\theta_{BN}+ \kappa_{\perp} sin^2\theta_{BN}.  \label{eq:totalKappa}
\end{equation} }
The maximum particle energy at the shock front decreases with increasing $r$ as the shock weakens when it
propagates out. The green curve is case II (the base case) which is for our Sun. The red curve is case I and corresponds to a star with a rotation speed half that of the Sun. We can see that the maximum energy for case I decreases faster than the base case. At 1 AU, the maximum particle energy at the shock front for the base case is $\sim 40$ MeV/nuc, but is  $< 30$ MeV/nuc for case I.
This is because the shock in the base case has an obliquity angle larger than case I and it leads to a smaller total $\kappa$ at the shock front and from equation~(\ref{tdyn}), we see that it corresponds to a higher maximum energy. For other cases (case III to case VI), the $E_{max}$  shows an interesting behavior: it quickly drops within $\sim 0.4$ AU, followed by a plateau-like period where the $E_{max}$ almost maintains a constant, and then it drops again.  The shock nose locations for these drops are:
$0.74$ AU for case VI, $0.88$ AU for case V, $1.10$ AU for case IV, and $1.47$ AU for case III.
 These locations are marked by the arrows in the left panel of Figure~\ref{fig:fig2}.
 It is interesting to notice that the $\theta_{BN}$'s at these locations are similar. As shown in the right panel of Figure~\ref{fig:fig2}, these turning points of $E_{max}$ correspond to a critical obliquity angle $\theta_{BN}=62^{\circ}$.
  One can understand this drop as the following: the  total diffusion coefficient, as given by equation~\ref{eq:totalKappa}, is
 dominated by the parallel diffusion coefficient $\kappa_{||}$ when $\theta_{BN} < 62^\circ$, and by the perpendicular $\kappa_{\perp}$
 when $\theta_{BN} > 62^\circ$. This difference leads to a ``break" of $E_{max}$ as a function of $r$.
  This can be so because $cos \theta_{BN}$ decreases with increasing $\theta_{BN}$, so the contribution of
 $\kappa_{||}$ to the total $\kappa$ decreases with  increasing $\theta_{BN}$. It has to be noted that
  in our model, $\kappa_{\perp}$ is decided by $\kappa_{||}$ through the NLGC theory and $\kappa_{||}$ is coupled to the excited waves at the shock front which  in turn is decided by the injection efficiency that depends on $\theta_{\perp}$ itself.
  How $\kappa$ vary with $\theta_{BN}$ is therefore non-linear and it is not straightforward to see how $\kappa_{||}$ or $\kappa_{\perp}$ dominates $\kappa$ at different values of $\theta_{BN}$. {Because this nonlinear dependence, one should not overemphasize the importance of a critial value
    of $62^{\circ}$ for $\theta_{BN}$. Using a different theory other than NLGC in describing $\kappa_{BN}$, the bevavior shown  in
    Figure~\ref{fig:fig2} may vanish. Indeed, the drop of maximum energy for case III (the blue curve) at $r=1.47$ AU is not as rapid as cases IV, V and VI. }

From the left panel of Figure~\ref{fig:fig2}, we see that at $0.7$ AU, $E_{max}$ for case VI is the largest; however,
at $1$ AU, $E_{max}$ for case IV and V are similar and are higher than other cases. At $r=1.5$ AU, $E_{max}$ for case III and IV are similar and are higher than other cases. This of course is the direct consequence of the ``break'' point occurring at different $r$'s for different cases.

%
%

Figure \ref{fig:fig2} suggests that the stellar rotation rate can affect planetary radiation environments via
the fluence of high energy particles.  The maximum energy acquired by the accelerated particles depends on a number of parameters including shock speed, shock compression ratio and shock geometry. From Figure \ref{fig:fig2} we can see that the particle maximum energy is greatest for solar-like stars with high rotation rates and decreases with the rotation rate. This has an impact of exoplanetary atmospheric environments.
{ For the terrestrial type exoplanet with Earth-like (1 bar) atmospheric pressure, particles with the energy of 1 GeV/nuc can penetrate to $\sim 3$  km from the ground, while particles at the low end of energies (10 MeV/nuc) will be stopped via collisions with atmospheric species at $70$ km, and thus pose little effect on chemistry or dosage of ionizing radiation.
The high energy particles (with energies $>$ 0.3 GeV) can penetrate into the lower planetary atmospheres (stratosphere-troposphere region) and induce enhanced ionization required to ignite atmospheric chemistry producing biologically relevant molecules  \citep{Airapetian2019}. Moreover, the strength of the planetary magnetic field is also important since the impact of SEP events depends on the fraction of the open magnetospheric field, which can support an efficient penetration of energetic particles into the planetary atmosphere. This fraction of the open field is moderated by the dynamic pressure of the stellar wind and CMEs associated with SEP events.
  The exoplanets with weaker magnetic fields and lower surface pressure will be more sensitive to lower energy particle penetration.}

\textcolor{black}{
  In passing, we note that although large SEP events are the greatest hazard during solar maximum,
  extended exposure to galactic cosmic rays is a greater hazard than SEPs during solar minimum.
  Indeed, work by \citet{Mewaldt2007} has shown that during solar minimum, fluence of GCR ions dominates that of SEP. }

\begin{figure}[htb]
   \centering
   \includegraphics[width=18.2cm, angle=0]{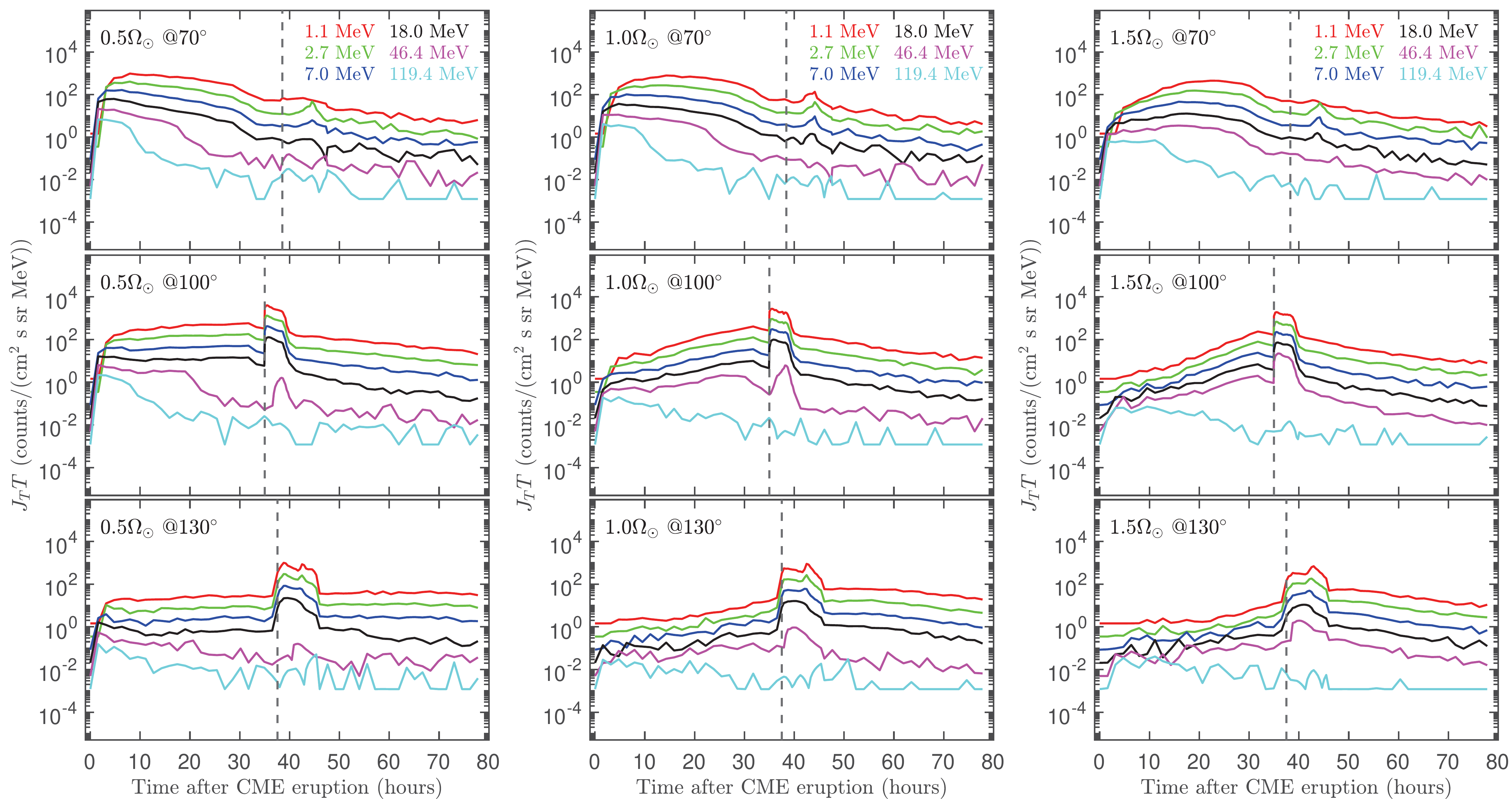}
   \caption{Time intensity profiles for case I, II and III. The vertical dash line in each panel marks the shock arrival time for that reference point.}
   \label{fig:fig3}
   \end{figure}

\begin{figure}[htb]
   \centering
   \includegraphics[width=18.2cm, angle=0]{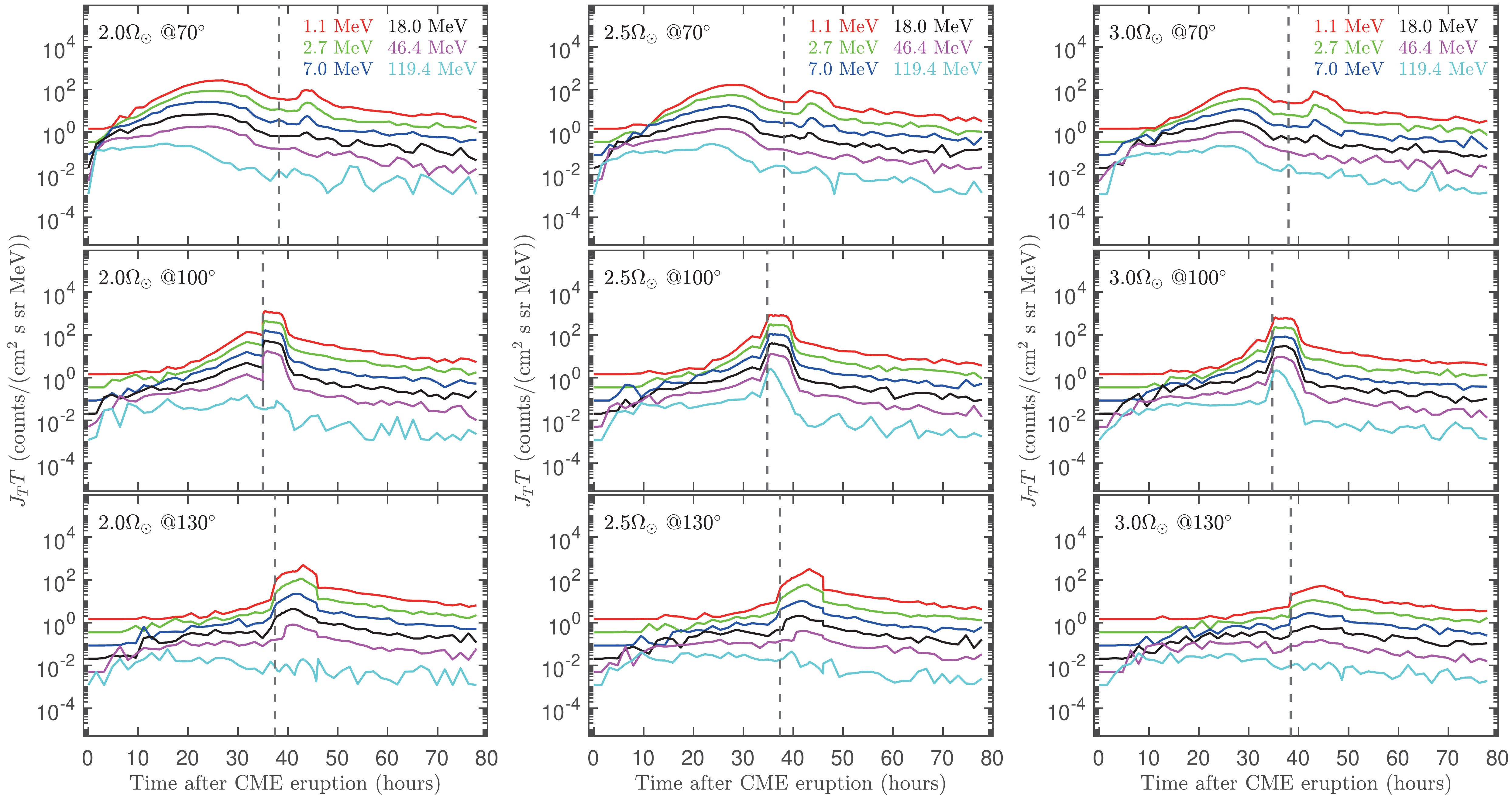}
   \caption{Same as Figure \ref{fig:fig3} but for case IV, V and VI.}
   \label{fig:fig4}
   \end{figure}

   To better understand the radiation environment at the Earth or other planets, time intensity profiles and particle fluence are
   necessary.  Figure \ref{fig:fig3} plots the time intensity profiles for case I, II and III.
   { The top panels are for observer A, located at  longitude $70^\circ$; the middle panels are for observer B, located at longitude $100^\circ$;
     the bottom panels are for observer C, located at longitude $130^\circ$.} Six energy channels are selected. The vertical dashed line marks the shock arrival.
   Figure \ref{fig:fig4} is the same as Figure \ref{fig:fig3}, but for cases IV, V and VI, respectively.
          The time intensity profiles for the same locations but with different rotation rates differ considerably.
          At point A ($\phi=70^\circ$), for all energies, we can see a fast rise followed by a gradual decay for case I. This is because the shock nose is
          magnetically connected to the observer at an early time but the connection moves to the eastern flank of the shock as the shock propagates out.
          For other cases, the rotation rates are faster, so the observers are initially connected to the western flank of the shock.
          The connections to the shock nose occur at $\sim$5.0, 16.7, 20.6, 23.0, and 24.6 hours for case II, III, IV, V, and VI, respectively.
          The rising phase for case III, IV, V, and VI are much slower than case I and a plateau-like period develops. As the shock further propagating out,
          the connections gradually move to the eastern flank and the intensity profiles begin to decay. 
          In all cases, no clear signatures of ESP phase occur at the shock passage.

          For point B ($\phi=100^\circ$), the time intensity profiles for all cases are increasing before the shock passes. This is because all observers are connected to the western flank of the shock and gradually move eastward. However, they never connect to the nose of the shock until the shock arrival.
          Different with point A, the ESP phases are very clear in all cases now, and the shapes of the ESP enhancement slightly differ for different rotation rates.
          After the ESP phases, the intensity profiles begin to decay. For point C ($\phi=130^\circ$), the observers are connected further to the western flank of the shock and they never connect to the shock nose before shock arrival. 
So the initial rises are shallower than point B.  After the shock passes, they begin to decay, but for low energy channels ($1.1$ MeV to $7.0$ MeV), they do not decay to a level lower than that before the shock arrival. This is because they are connected to the shock nose where these lower energy particles are trapped with the shock after the shock passage.
The strong longitudinal dependence of the time-intensity profile is the same as reported in \citet{Hu2017,Hu2018}. The inclusion of the ESP phase and its longitudinal dependence is a new feature of the updated iPATH model.

\begin{figure}[htb]
   \centering
   \includegraphics[width=16cm, angle=0]{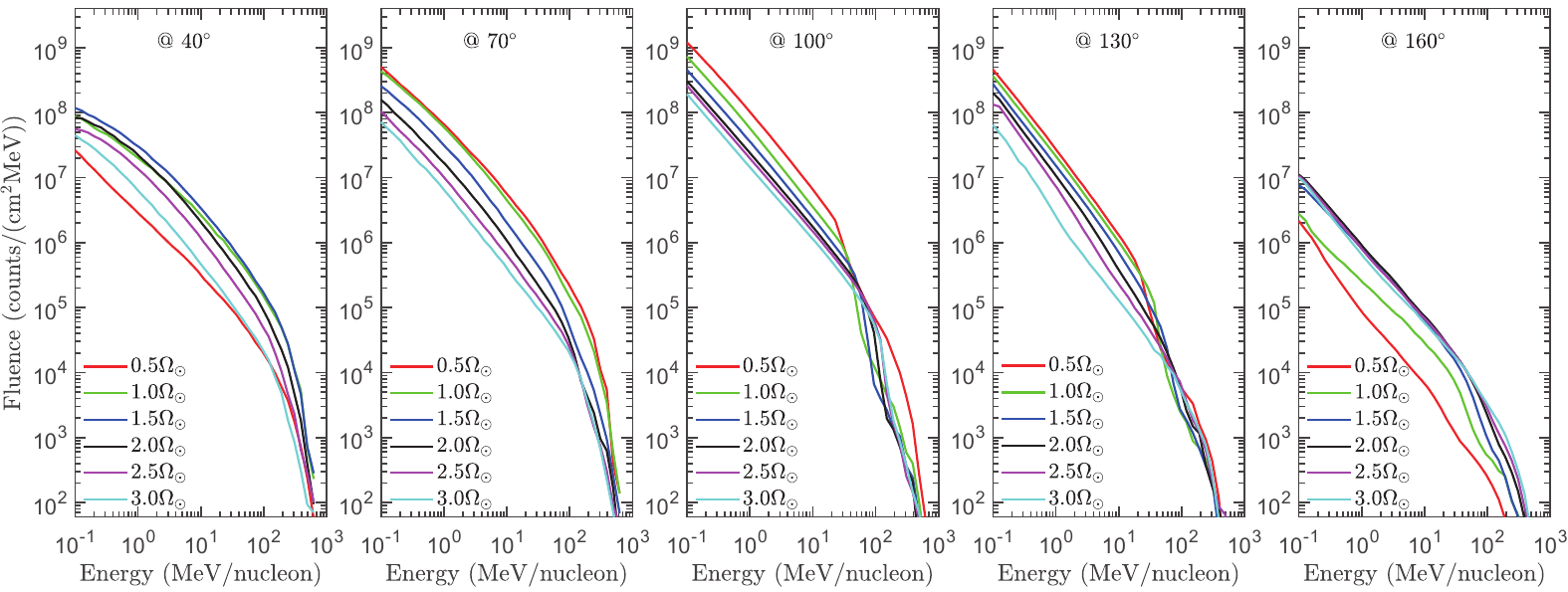}
   \caption{Event-integrated spectra at longitudes of $40^\circ$, $70^\circ$, $100^\circ$, $130^\circ$, and $160^\circ$, respectively.}
   \label{fig:fig5}
   \end{figure}

   Figure~\ref{fig:fig5} plots the event-integrated spectra at 1 AU for $5$ locations. From left to right, these correspond to observers at
     $\phi$=40$^\circ$, $\phi$=70$^\circ$, $\phi$=100$^\circ$, $\phi$=130$^\circ$, and $\phi$=160$^\circ$. Comparing to the time intensity profiles shown in
     Figure~\ref{fig:fig3} and Figure~\ref{fig:fig4}, two more observers, at  $\phi$=40$^\circ$ and at  $\phi$=160$^\circ$ are added.
     It can be seen that both the magnitudes and the shapes of the event-integrated spectra vary with observer's longitude and star's rotation rate.
     Note that for all observers, as the rotation rate of the star increases, the connection to the shock shifts from right to left.
     This means an observer at a particular location may not be magnetically connecting to the shock for the whole event.
     For the observer at $\phi$=40$^\circ$,   the fluence for the $0.5 \Omega_0$ case is the smallest. This is because at the beginning of the event
     the connection to the shock is already at the right flank. As the star rotation rate increses to $1.0 \Omega_0$, the connection (at any give time) moves to the left  along the shock front,  therefore being able
     to access the shock nose at earlier times where the acceleration is the strongest. When the star rate further increases, however, the connection at early times
     shifts to the left flank of the shock, where the acceleration is weaker than the case of $1.0 \Omega_0$. Consequently the fluence decreases.
     For the observer at  $\phi$=70$^\circ$,  the fluence for the $0.5 \Omega_0$ and  $1.0 \Omega_0$  cases are similar and are the largest. Again, this is because
     in these two cases, the observer has the best connection to the shock (longer duration) comparing to other cases.
     For the observer at  $\phi$=100$^\circ$, the event is a central Meridian event (since the CME propagates toward $\phi$=100$^\circ$).
       Comparing to other {longitudes}, the fluence at low energies is the largest at  $\phi$=100$^\circ$. Except the  $0.5 \Omega_0$ case, the variation at high energies
       for different $\Omega$s are small. This is also true for  the $\phi$=130$^\circ$ cases, where at high energies the fluence for different rotation rates are similar.
     This is because high energy particles are accelerated close to the shock, and for the observer at
     $\phi$=130$^\circ$,the initial connection to the shock is to the left flank of the shock for all rotation rates, and the accelerated particle spectra at a
     broad range of the flank do not vary much. The fluence seen at $\phi=160^{\circ}$ is interesting. First of all, the fluences for all rotation rates are
     smaller than those at other $\phi$'s. This is because the observer  at $\phi=160^{\circ}$ only connects magnetically to the shock after the shock propagates
     beyond 1 AU. Consequently almost all particles (besides some that diffuse cross fields) observed at $\phi=160^{\circ}$ propagate inward from the shock when
     the shock is beyond 1 AU. A
     faster rotation means that the observer connects to the shock nose at an earlier time. Therefore the fluence (especially at higher energies) is
     larger for dtars with greater rotation rates, as shown in the figure.
     For the observer locates at the nose of the shock (i.e., $\phi$=100$^\circ$),  clear spectral breaks can be seen. The spectral break energy is related to the
     maximum energy at the observer location.  This can be seen by examining the right panel of Figure \ref{fig:fig2} with the middle panel of Figure \ref{fig:fig5}
     as they correspond to the same longitude. Consider the red curve (case I), the green curve (case II), and  the cyan curve (case VI) as examples, their break energies
     are $\sim$29.0, $\sim$38.0, and $\sim$90.0 MeV, and the corresponding maximum energies at 1 AU are $\sim$26.7, $\sim$43.2, and $\sim$92.5 MeV, respectively.

\section{Conclusion and Discussion} \label{sec:intro}


In this letter, we report the results of numerical simulations to examine the effect of star’s rotation rate on the properties of energetic particles events using the 2D iPATH model. We consider $6$ scenarios for stellar rotation rates ranging from $0.5 \Omega_{\sun}$ to $3.0 \Omega_{\sun}$. Maximum particle energy along the shock front is derived for these scenarios. We also model the time intensity profiles  at three locations ($\phi=70^\circ,100^\circ,130^\circ$) and particle spectra at five locations ($\phi=40^\circ$, $70^\circ$, $100^\circ$, $130^\circ$, $160^{\circ}$). Our results show that the characteristics of SEP events can be
affected by the star rotation rate. { The rotation rate changes the magnetic connection of the observer to the shock, and  alters the shock geometry.
  The maximum energy gained by a particle at the shock front depends on the shock geometry.
  The observed time intensity profiles and event integrated spectra also depend on the
  star's rotation rate. However, these dependences on the rotation rate are less prominent as to the maximum energy at the shock.
  This is due to the fact that the shock front is spatially extended and any given observer over a certain period of time will probe a range of the shock front, leading to a time
  intensity profile that is less sensitive to both the longitude and star rotation rate. Since our results show that at high energies the effect of star rotation on the fluence is
  not significant,  it suggests that the variation of star rotation has limited impact on atmospheric chemistry of exoplanets with high surface pressure.
  However, as discussed above, for atmospheric pressures less than $0.5$ bar, the impact of SEPs with energies $<$ 100 MeV/nuc becomes significant for prebiotic
  chemistry (Airapetian et al. 2019).
  On the other hand, the fluence of  lower energy particles have stronger dependence on magnetic field geometry, and therefore the general radiation environment is affected by star's rotation.  }

 Our results provide the first insights for the SEP properties in response to various magnetic field geometry. They form a framework to study stellar SEP events including their fluence, spectra and maximum energy from stars of different magnetic activity levels that vary over the course of stellar evolution. Higher level of magnetic activity results in denser and strongly magnetized corona that will drive more energetic flares and CME events. {Such energetic CMEs can drive shocks lower in the stellar corona than that on the current Sun \citep{Lynch2019, Airapetian2019}. Our future studies of  stellar SEP event will use the results of data driven magnetohydrodynamic models of evolving solar-like stars based on multi-observatory coordinated observing campaign involving NASA's Transiting Exoplanets Survey Satellite (TESS), Hubble Space Telescope, HST, X-ray missions XMM-Newton, NICE and ground-based spectropolarimetry (\citep{Airapetian2019}).} These studies will be useful in estimating the radiation environment of other Earth-like exoplanets orbiting solar-like star. This is especially important in light of the upcoming James Webb Space Telescope (JWST) that will characterize atmospheric chemistry of exoplanets in the solar neighborhood.



\acknowledgments

This work is supported at the University of Alabama in Huntsville under NASA grants NNX17AI17G, NNX17AK25G, and 80NSSC19K0075. S.F. and Y.J. acknowledge partial support of National Key R\&D Program of China (2018YFC1407304, 2018YFF01013706),  Open Fund of Key Laboratory (201801003), and other Foundation (315030409). V.S.A. is supported by NASA grant 80NSSC17K0463, TESS Cycle 1 grant 80NSSC19K0381 and GSFC ISFM SEEC grant. S.F. and G.L. acknowledge China University of Geosciences (Beijing) for providing fast and adequate computing resources.

\end{document}